\documentclass[aps,showpacs,amsmath,amssymb,preprint]{revtex4}
 \usepackage{graphicx}
 \usepackage[colorlinks=true]{hyperref}

 \def\T{\textstyle}
 \def\l{\left}
 \def\r{\right}
 \def\nf{n_{\!f}}

 \def\be{\begin{equation}}
 \def\ee{\end{equation}}
 \def\bea{\begin{eqnarray}}
 \def\eea{\end{eqnarray}}
 \def\bean{\begin{eqnarray*}}
 \def\eean{\end{eqnarray*}}
 \def\esim{\mathrel{\rlap{\lower0.2em\hbox{$-$}}\raise0.2em\hbox{$\sim$}}}
 \def\gsim{\mathrel{\rlap{\lower0.2em\hbox{$\sim$}}\raise0.2em\hbox{$>$}}}
 \def\ksim{\mathrel{\rlap{\lower0.2em\hbox{$\sim$}}\raise0.2em\hbox{$<$}}}
\begin{document}

\title{Collisional jet quenching becomes probable}
\author{A.~Peshier}
\affiliation{
 Institut f\"{u}r Theoretische Physik,
 Universit\"{a}t Giessen, 35392 Giessen, Germany}
\date{\today}

\begin{abstract}
It was argued recently that loop corrections to tree-level amplitudes are essential in the discussion of the collisional energy loss of energetic partons in the hot quark gluon plasma: Instead of $dE_{\rm coll}^B/dx \sim \alpha^2 T^2 \ln(ET/m_D^2)$, as first derived by Bjorken (assuming a constant $\alpha$), the mean energy loss actually behaves as $dE_{\rm coll}/dx \sim \alpha(m_D^2)T^2$.
Here we calculate, within this resummation-improved framework, the probability distribution functions (`quenching weights') of the collisional energy loss.
First results from a Monte Carlo implementation of this probabilistic collisional quenching shed new light on the interpretation of jet suppression in heavy ion collisions.
\end{abstract}


\maketitle

\section{Introduction}
The dominance of radiative energy loss \cite{Baier:1996kr,Zakharov:1996fv,Gyulassy:2000fs} for the phenomenon of jet quenching in heavy ion collisions seemed long established when Mustafa and Thoma \cite{Mustafa03} re-considered, in 2003, the possibility of collisional partonic quenching \cite{Bjorken:1982tu,TG,Braaten:1991jj}.
However, only recently, after publication of new data on the nuclear suppression factor from non-photonic electrons \cite{Phenix}, this idea seems to have gained wider interest, see e.\,g.\ \cite{Gyulassy05}.
In fact, an additional contribution to the partonic energy loss appears necessary to reconcile data-adjusted model parameters with general expectations, say in the GLV formalism \cite{Gyulassy:2000fs} the partonic rapidity density $dN/dy$ with the final entropy density, cf.\ \cite{Muller:2005wi}. Also the transport coefficients $\hat q$ inferred in \cite{Dainese:2004te,Eskola:2004cr} within the BDMPS-Z formalism \cite{Baier:1996kr,Zakharov:1996fv} using probabilistic `quenching weights' \cite{Salgado:2003gb} are almost an order of magnitude larger than estimated \cite{Baier03}.
In this situation, a collisional component of the energy loss is a welcomed remedy, as advocated in \cite{Gyulassy05}.

These (as well as previous) considerations rely on Bjorken-type formulae for the (mean) collisional energy loss, $dE_{\rm coll}^B/dx \sim \alpha^2 T^2 \ln(ET/m_D^2)$. However, as pointed out recently \cite{Peshier:2006hi}, the collisional energy loss is an observable for which loop corrections to the tree level approximation are essential. Taking these higher order terms into consideration by using the {\em running} coupling in the calculation leads to modified predictions, which differ even parametrically from the commonly used expressions, see Eq.~(\ref{eq.DeltaE}) below.

In this note, in Section \ref{sec.dEdx}, we will briefly outline the main idea of the (re-)calculation of the mean energy loss. In Section \ref{sec.PDF}, the probability distribution functions for the collisional energy loss are derived, which are necessary to describe jet quenching. These findings are implemented in a Monte Carlo model which will be presented, along with results, in Section \ref{sec.PCQ}.
The implications of these exemplary studies with regard to heavy ion phenomenology are summarized in the Conclusions.

\section{Mean collisional energy loss \label{sec.dEdx}}

Following Bjorken \cite{Bjorken:1982tu}, the mean collisional energy loss of an energetic parton ($E \gg T$) in a hot QGP ($T \gg \Lambda$) can be calculated from its weighted collision rate, which is determined by the flux and the cross section,
\be
	\Delta E_{\rm coll}^j
	=
	\sum_s
	\int_{k^3} \rho_s(k)\, \Phi 
	\int dt\, \frac{d\sigma_s^j}{dt}\, \omega \, \Delta x \, .
\label{eq.dEdx}
\ee
The energy transfer in one collision, $\omega = E-E'$, is related to the angle between the `jet' $j$ and the scatterer $s$ as well as the invariant momentum transfer by $t = -2(1-\cos\theta)k\omega$.
Furthermore, $\Phi = 1-\cos\theta$, and $\rho_s = d_s n_s$ is the density of the scattering partons, with $d_g = 16$ and $d_q = 12\nf$ for gluons and quarks, and $n_\pm(k) = \l[ \exp(k/T) \pm 1 \r]^{-1}$ in the ideal gas approximation.

Since processes with small momentum exchange dominate, the cross sections can be approximated by the $t$-channel contribution, $d\sigma/dt \sim \alpha^2/t^2$. In strictly tree-level approximation, $\alpha$ is a {\em constant} parameter. Thus,  the predictive power of the resulting expression is questionable because $\Delta E_{\rm coll}$ can probe a large range of the virtuality $t$ where the running of the coupling, as a result of loop corrections, is important. On the other hand, these quantum corrections are also formally required to screen the long-range interactions in the medium.

As recollected in \cite{Peshier:2006hi}, in a first approximation the loop corrections can be taken into account by evaluating the differential cross section with the running coupling
\be
	\alpha(t)
	=
	\frac{b_0^{-1}}{\ln(|t|/\Lambda_{QCD}^2)} \, ,
\label{eq.alpha(t)}
\ee
where $b_0 = (11 - \frac23\, \nf)/(4\pi)$, and by imposing an IR cut-off in $t$, $\mu^2 \sim m_D^2$, related to the Debye mass.
Only the latter aspect has been utilized in the past, which led to Bjorken's formula \cite{Bjorken:1982tu} as well as various refinements thereof, which aimed at the precise determination of the cut-offs \cite{TG,Braaten:1991jj}.
However, as pointed out in \cite{Peshier:2006hi}, taking into account the running coupling modifies even the {\em structure} of the result. The $t$-integral in (\ref{eq.dEdx}), with $\omega \sim t$, is basically $\int dt/(t\ln^2 t) = -1/\ln t$ vs.\ $\int dt/t = \ln t$ in the case of a constant coupling, hence the perturbative order and the dependence on the cut-offs is changed completely. The UV cut-off being related to the maximal possible energy transfer, $\omega_{\rm max} = {\cal O}(E)$, the mean collisional energy loss becomes $E$-independent for infinite parton energy \cite{Peshier:2006hi},
\be
	\Delta E_{\rm coll}
	\sim
	\alpha(\mu^2) T^2 \Delta x \, ,
\label{eq.DeltaE}
\ee
which is the direct imprint of asymptotic freedom. For a more detailed discussion of the relation of Eq.~(\ref{eq.DeltaE}) to Bjorken's formula we refer to \cite{APnew}.

It was argued in \cite{Peshier:2006hi} that such an `improved' perturbative framework might be useful to estimate the energy loss even for rather small temperatures near $T_c$ as relevant in heavy ion phenomenology. The main idea is to realize that the energy loss probes a (complicated) QCD 4-point function in the special limit $-t \ll s$. The same 4-point function determines also $(i)$ the distance dependence of the strong coupling as obtained from the heavy-quark potential at $T = 0$, and $(ii)$ the screening of this potential at $T \ge T_c$, as encoded in the Debye mass. Now both observables have been calculated in the large-coupling regime within lattice QCD. Remarkably, these non-perturbatively results -- for both quantities -- are reproduced {\em quantitatively} by the resummation-improved perturbation theory \cite{Peshier:2006ah} with only one parameter, whose adjustment yields, for $\nf = 2$, the canonical value $\Lambda_{QCD} = 0.2$GeV. This unexpected, but pleasing observation and the close relation of the three quantities may then support also an (otherwise not justifiable) extrapolation in case of the energy loss. 

In the revised form, the collisional energy loss is larger than previous estimates based on Bjorken-type formulae, e.\,g.\ $\Delta E_q/\Delta x \approx 3$GeV/fm for $T = 3T_c \approx 0.5$GeV. It is emphasized again that, after fixing $\Lambda_{QCD}$ from lattice QCD data, these results are basically parameter-free (while previous estimates had to `guess' a value for the {\em fixed} parameter $\alpha$.).

\section{Collision probability distribution function \label{sec.PDF}}

While the estimates above already hint at the importance of collisions for jet quenching, the {\em mean} energy loss is too crude to quantify the effect in heavy ion collisions. The fact that the initial hard parton spectrum is steeply decreasing requires to consider the probability density function \cite{Baier}, which in the context of the radiative energy loss is sometimes dubbed quenching weights \cite{Salgado:2003gb}.

For the collisional energy loss mechanism studied here, the probability density for an energy transfer $\omega$ to the medium along the path $\Delta x$ reads \be
	P_{\Delta x}^j(\omega; T, E)
 	=
	\sum_s d_s
	\int_{k^3} n_s(k) \Phi
	\int dt\, \frac{d\sigma_s^j}{dt}\, \delta(\omega-\omega(t)) \, \Delta x \, .
\label{eq.P(omega)}
\ee
Note that for larger $\Delta x$ it is rather to be interpreted as the average number of collisions with an energy transfer $\omega$.
Obviously, its first moment $\int d\omega P_{\Delta x}^j(\omega; T, E)$ leads back to the mean energy loss (\ref{eq.dEdx}).

Within the same calculation framework as outlined in the previous Section (for more details see \cite{Peshier:2006hi}), the $t$-integral gives
\[
	2(1-\cos\theta) k \l. \frac{d\sigma}{dt} \r|_{t^\star}
\]
with the implicit constraint that $|t^\star| = 2(1-\cos\theta)k\omega$ falls into the integration interval $[\mu^2,2(1-\cos\theta)k\omega_{\rm max}]$. In the following we set $\omega_{\rm max} = E$ 
\footnote{This can be justified {\em a posteriori}.}; 
then $P_{\Delta x}^j(\omega; T, E)$ vanishes at $\omega = E$ (and at higher energies).

\begin{figure}[ht]
 \centerline{\includegraphics[width=12cm]{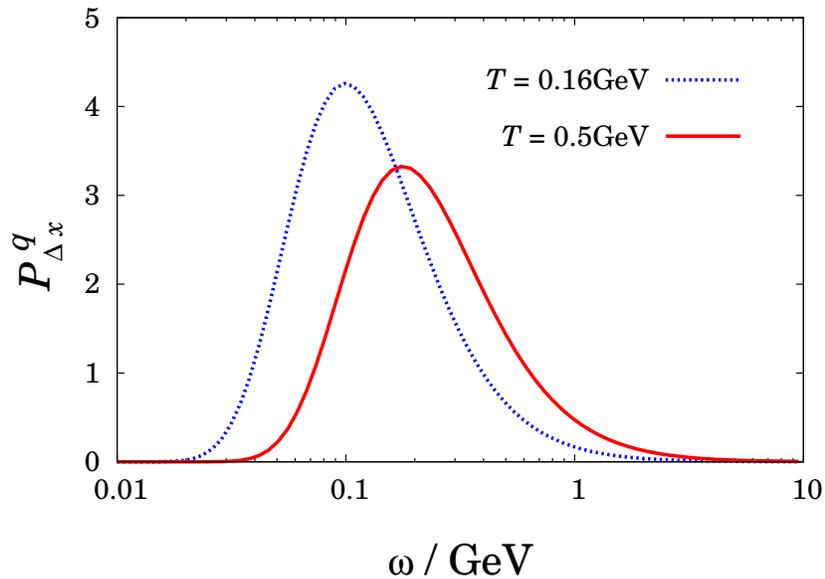}}
 \caption{The probability density function (\ref{eq.Pqg(omega)}) for collisions with an energy transfer $\omega$, evaluated here for quarks, $\Delta x = 1$fm  and two representative temperatures.
 	\label{fig.PDF}}
\end{figure}

For the cross sections `improved' by the running coupling in the form (\ref{eq.alpha(t)}), the angular integral in Eq.~(\ref{eq.P(omega)}) can be performed analytically and one readily arrives, without further approximations, at
\be
	P_{\Delta x}^{q,g}(\omega; T)
 	=
	\l( \T\frac23 \r)^{\!\pm 1}
	\frac3{\pi b_0^2} \frac{T \Lambda^2}{\omega^3}
	\l( {\cal I}_+ + \frac\nf3 {\cal I}_- \r)
\label{eq.Pqg(omega)}
\ee
for quarks and gluons, respectively.
The dimensionless functions are defined by
\[
	{\cal I}_\pm(a,b)
	=
	\int_{a/b}^\infty d\zeta\, \frac{I(a)-I(\zeta b)}{\exp(\zeta) \pm 1} \, ,
\]
with $I(x) = x/\ln(x)-{\rm li}(x)$, where ${\rm li}(x)$ denotes the logarithmic integral function. In Eq.~(\ref{eq.Pqg(omega)}), ${\cal I}_\pm(a,b)$ are evaluated for $a = \mu^2/\Lambda^2$ and $b = 4T\omega/\Lambda^2$.
In the following we will use $\mu^2=\frac12\, m_D^2$ \cite{Peshier:2006hi}, and determine the Debye mass self-consistently \cite{Peshier:2006ah}
\be
	m_D^2 = \l( 1+\T\frac16 \nf \r) 4\pi\alpha(m_D^2)\, T^2 \, .
\label{eq.mD}
\ee
As depicted in Fig.~\ref{fig.PDF}, $P_{\Delta x}^{q,g}(\omega; T)$ is, for typical temperatures, peaked at ${\cal O}(0.1)$GeV.
It is comforting to note that it decreases rapidly for larger values of $\omega$, which guarantees that the specific choice of $\omega_{\rm max}$ is unimportant, as anticipated above. For increasing temperatures, the distribution shifts to larger $\omega$ and it becomes wider.

\begin{figure}[ht]
 \centerline{\includegraphics[width=12cm]{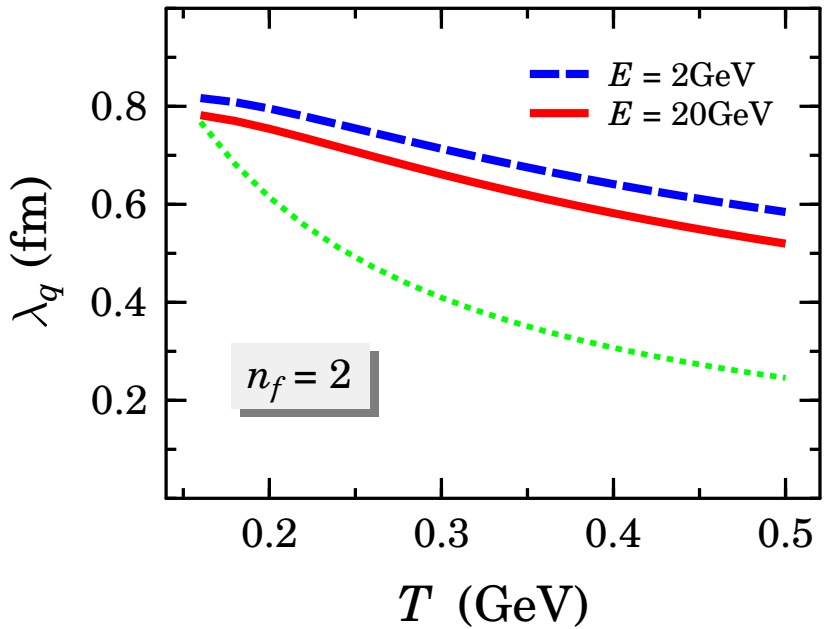}}
 \caption{The quark mean free path as a function of the temperature, for two different quark energies. Shown by the dotted line for comparison is the estimate of the typical particle distance $r_0$ in the medium.
 	\label{fig.lambda}}
\end{figure}

For sufficiently small $\Delta x$, when $P_{\Delta x}(\omega)$ is a probability distribution in the literal sense, the expression $1-\int d\omega P_{\Delta x}(\omega;T) > 0$ is the chance that no collision happens within $\Delta x$. Accordingly, the zeroth moment of the probability distribution defines the mean free path $\lambda(T;E)$ of the hard parton in the medium,
\be
	\int d\omega P_\lambda(\omega; T) = 1 \, .
\ee
It is instructive to estimate the size of the mean free length as a phenomenologically important quantity.
First, one can verify again that predictions of the present approach are not very sensitive to the hard momentum scale; the mean free path lengths of a parton with intermediate energy differs from the asymptotic limit $E \rightarrow \infty$ only moderately, see Fig.~\ref{fig.lambda}. 
Second, $\lambda$ is of the same order of magnitude as the particle distance, which can be estimated from the density $\rho_0 = \sum_s d_s n_s$. This is a clear sign of the strong coupling in the medium (sQGP). While the quark free length coincides with $r_0 = \rho_0^{-1/3} \sim T^{-1}$ at about $T_c$, the gluon mean free path, due to the larger color charge, becomes smaller than $r_0(T)$ already at $\hat T \approx 0.5$GeV. Interestingly, $\hat T \approx 4T_c$ is close to the value estimated, by other means, in \cite{Peshier:2005pp} for the `transition' from a strong-coupling to a moderate-coupling regime.

\section{Probabilistic collisional quenching \label{sec.PCQ}}
To quantify the effect of collisional energy loss, we develop a dynamical model of jet quenching in heavy ion collisions. Given the complexity of the phenomenon, which results in theoretical uncertainties on various levels, the intention is, for now, not to formulate the dynamics as realistically as possible; we rather focus on the transparency of the model. Having only few parameters, which hopefully capture essential aspects, might help to constrain the theoretical uncertainties in the underlying energy loss mechanisms.

We consider here only the partonic energy loss due to collision within the dense plasma above $T_c$, i.\,e., we explicitly `switch off' radiation and we do not account for effects at and after hadronization. The first assumption is, of course, unrealistic but will provide for the benchmark we are aiming at. The second assumption, on the other hand, relies on the idea that a parton fragments only after it has left the medium.

For the sake of simplicity, we study central collisions and focus on the mid-rapidity region. We assume a boost-invariant Bjorken expansion, $T(\tau) = T_0 (\tau/\tau_0)^{-1/3}$ \footnote{This simple parameterization might actually 
	be not too unrealistic -- while a radial expansion accelerates the cooling, 
	this effect might be counterbalanced by a non-ideal equation of state and a 
	non-ideal hydrodynamics.},
with a homogeneous transverse profile. Consequently, the radius $R$ of the longitudinally expanding cylinder is an {\em effective}  parameter, say $R \approx 5$fm \footnote{This will be a trade-off between the nuclear skin and the 
	subsequent radial expansion, which we will disregard here.}.
The formation time of the QGP is related to the saturation scale, for RHIC $\tau_0 \approx 0.2$fm is a generic value in the literature \footnote{To 
	simplify matters we ignore a less important difference in the formation 
	times of the medium and the hard partons.}.
The initial temperature $T_0$, as the third parameter, is related to the total energy/entropy of the system; $T_0 \approx 0.5$GeV is a representative value.
To test the sensitivity of predictions on these parameters for the dynamics/geometry of the collision, they will be varied from the above values, as summarized in Table 1.
\begin{table}[ht]
\begin{tabular}{c||c|c|c|c|c|c}
	&\ set 1 \ &  set 2 \ &\ set 3 \ &\ set 4 \	&\ set 5 \ 	&\ set 6\\ \hline
	$R$ (fm)		& 5		& 3		& 5		& 5		& 5	 	& 7		\\ \hline
	$T_0$ (GeV)		& 0.5	& 0.5 	& 0.3	& 0.7	& 0.5	& 0.5	\\ \hline
	$\tau_0$ (fm)	& 0.2	& 0.2	& 0.2	& 0.2	& 0.1	& 0.2	\\ \hline
\hline
	$\tau_c$ (fm)	& 6.1	& 6.1	& 1.3	& 16.7	& 3.1	& 6.1
\end{tabular}
\caption{The different sets of (effective) parameters to describe the collision. The last row gives $\tau_c = \tau_0 (T_0/T_c)^3$ for $T_c = 0.16$GeV, i.\,e.\ the QGP life time in the present approach.}
\label{tab.params}
\end{table}

The hard partons are initialized uniformly in the transverse plane (which adds to the fact that $R$ is to be considered an effective radius). Depending on the `jet vertex', characterized by $r \ (<R)$ and the emission angle $\varphi$ with respect to the radius vector, the partons propagate a length
\[
	l_\pm
	=
	\l( R^2-r^2\sin^2\varphi \r)^{1/2} \pm |r\cos\varphi|
\]
in the transverse plane (ignoring the small deflections by the collisions).
By taking the modulus of the second term, $l_+ > l_-$; below some results will be given separately for `far side' and `near side' jets.
To segregate phenomena like the Cronin effect, we consider larger transverse momenta, $p_t \gsim 4$GeV, where the initial parton spectrum is calculable in pQCD; here we adopt the parameterization
\[
	\frac{dN}{d^2p_t}
	\propto
	\l( 1+\frac{p_t}{p_0} \r)^{-\nu} \, ,
\]
with $p_0 \approx 1.75$GeV and $\nu \approx 8$ \cite{Mueller}.

In the course of time, as the hard partons propagate through the comoving medium, the $p_t$-spectrum evolves due to collisions with the thermal partons. The probability $P_{\Delta x}(\omega; T)$ of an energy transfers $\omega$ along the stretch of path $\Delta x$ depends on the {\em local} temperature of the rapidly cooling plasma.
This important fact is taken into consideration in a Monte-Carlo simulation by  discretizing the path such that $\Delta x$ is smaller than the typical time scale on which $P_{\Delta x}(\omega; T(\tau))$ changes. Simultaneously, to adhere to the probability interpretation of $P_{\Delta x}$, $\Delta x$ has to be smaller then the mean free length at the given time. For the implementation, $\Delta x = \frac13 \lambda$ is an appropriate choice between numerical accuracy and computational efficiency.

\begin{figure}[ht]
 \centerline{\includegraphics[width=12cm]{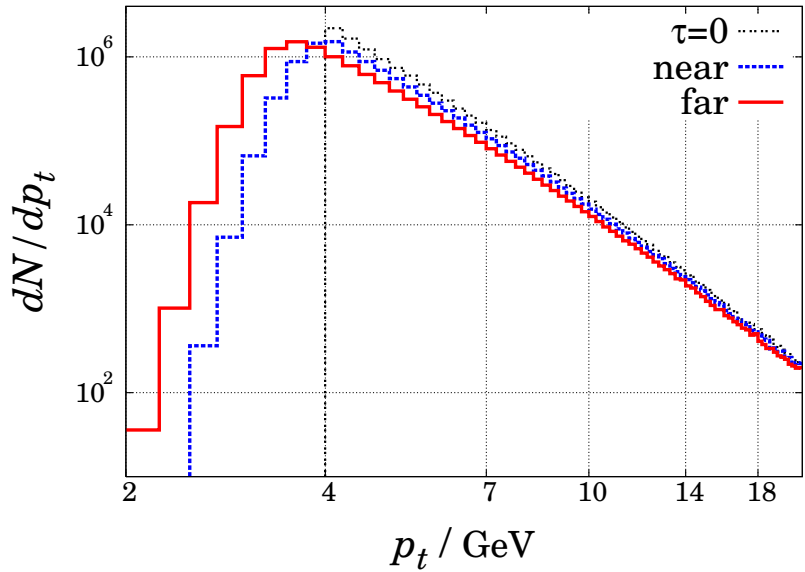}}
 \caption{The evolution of the hard parton spectrum from formation time to freeze out, plotted separately for `near-side' and `far-side' partons, for set 1 of model parameters, cf.\ Table~\ref{tab.params}.
 	\label{fig.Npt}}
\end{figure}

\begin{figure}[ht]
 \centerline{\includegraphics[width=12.2cm]{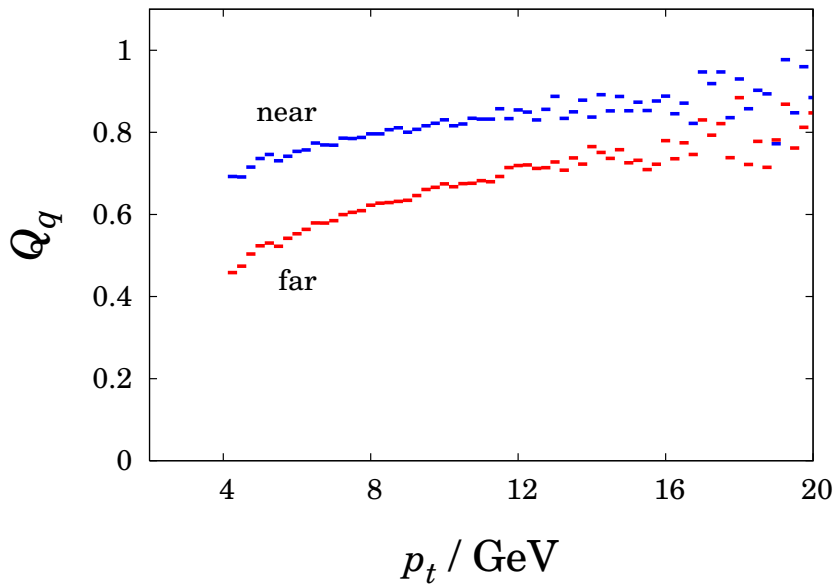}}
 \caption{The quenching factor corresponding to the spectrum in Fig.~\ref{fig.Npt}, given separately for `near-side' and `far-side'. For large $p_t$ events become rarer, which increases the statistical fluctuations.
\label{fig.Qq_set1}}
\end{figure}

Fig.~\ref{fig.Npt} shows the result of a simulation, for set 1 of parameters, with $10^7$ `jets'.
The spectrum shifts towards smaller $p_t$ -- although considerably different for the far and near side partons, as expected from comparing the life time, $\tau_c \approx 6$fm in this case, to the typical path lengths (obviously, $l_- < R$).  
These differences are better visible in the quenching ratio
\be
	Q_q
	=
	\frac{dN^{(\tau_c)}/dp_t}{dN^{(\tau=0)}/dp_t}
\ee
shown in Fig.~\ref{fig.Qq_set1}.
Although the present model is certainly too crude to make rigorous statements, such a distinct difference -- even for central collisions and with no transverse expansion taken into account -- is interesting with regard to recent results on jet correlations.
Another noteworthy fact is the increase of $Q_q$ with $p_t$ -- which is {\em qualitatively} different from previous predictions \cite{Mustafa03,Gyulassy05}. The reason is that the later approaches were based on Bjorken-type formulae where, on average, $dE_{\rm coll}^B \sim \ln E$.
However, as argued above, the dependence of the mean collisional loss on the parton energy is actually much milder: $dE_{\rm coll} \sim E^0$ for $E \rightarrow \infty$.

Finally, I briefly discuss the absolute size of the quenching factor. As discussed before, the collisional probability distribution function (\ref{eq.Pqg(omega)}) is basically parameter free, so it is first comforting that, for a reasonable geometry/dynamics, the predictions {\em without} radiative loss do not undershoot the observed {\em nuclear} quenching factor, $R_{AA} \sim 0.2$.
Moreover, with an estimated light quark radiative quenching factor $\sim 0.6$ \cite{Gyulassy05}, a collisional quenching $\sim 0.5$ (averaged over near and far side) seems just in the expected range \footnote{Note, however, that the 
	actual situation might be more involved than incoherent collisional and 
	radiative quenching, as recent considerations \cite{Peigne} 
	demonstrate.}. 
Since the radiative modification factor is almost constant over a large $p_t$-range \cite{Gyulassy05}, the moderate increase of $R_{AA}$ seen at large $p_t$ \cite{Phenix} could be a more indicative signature of the collisional contribution to the partonic energy loss.

Before concluding it is necessary to investigate the sensitivity of the results on the model parameters describing the geometry/dynamics of the collision.
\begin{figure}[ht]
 \centerline{\includegraphics[width=12cm]{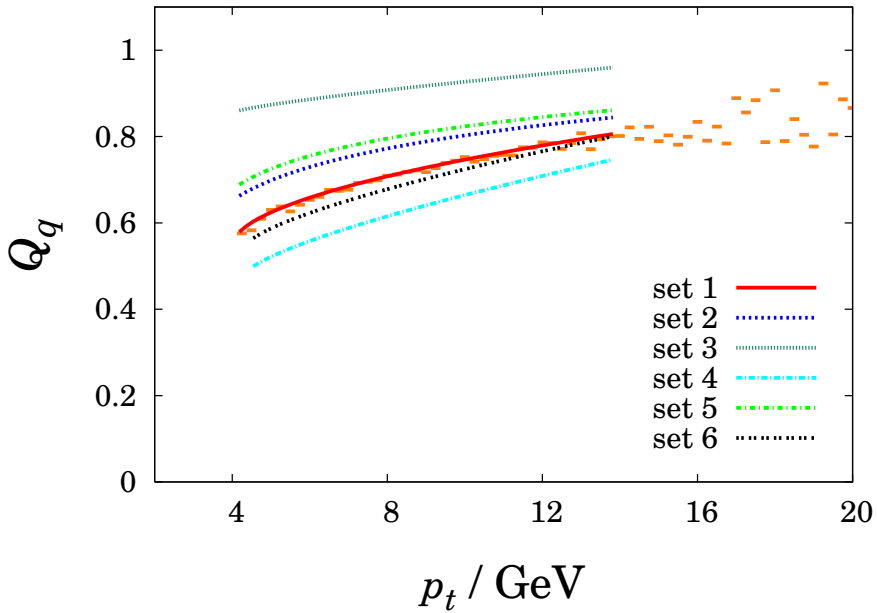}}
 \caption{The quark quenching factor for the various sets of parameters summarized here by simple fits to the Monte Carlo results (depicted only for set 1) for $p_t \in [4,14]$GeV (above 14GeV the statistical fluctuations become sizable).
 	\label{fig.Q_pt_compare}}
\end{figure}
From the results summarized in Fig.~\ref{fig.Q_pt_compare} it can be inferred that a very low initial temperature (set 3) leads to a diminished collisional quenching. A smaller effective radius (set 2) or a reduced life time due to a shorter formation time (set 5) give just a moderate change towards larger $Q_q$ values. A larger radius (set 6) only marginally decreases $Q_q$ compared to set 1, which is basically a consequence of the life time. On the other hand, larger initial temperatures give somewhat larger collisional quenching factors (set 4).

\section{Conclusions}
The fact that the average collisional energy loss per unit length has a different parametric behavior than widely held is reflected also in the corresponding probability density functions. In a resummation-improved framework, by taking into account the running coupling, these `collisional quenching weights' can be predicted unambiguously from the lattice-QCD adjustable value of $\Lambda_{QCD}$ -- without having to assume a value for the coupling strength.
Besides the mean energy loss, a second relevant moment of the probability distribution function is the mean free path of hard partons. Estimates for phenomenologically interesting temperatures yield the same order of magnitude, $\ksim 1$fm, as the typical particle distance in the QGP -- which is another quantitative indication of a strongly coupled system.

The probabilistic collisional energy loss can be readily implemented in a Monte Carlo code for jet quenching. A fully dynamical treatment of the quenching weights is feasible \footnote{By contrast, the models \cite{Dainese:2004te,Eskola:2004cr} rely on a scaling law \cite{baier98} relating the expanding medium to a static situation}, thus a more involved dynamics of the rapidly cooling system can, in principle, be accurately taken into account.

The present investigation is, in modeling the dynamics of heavy ion collisions, exemplary; more detailed results will be presented in a forthcoming study.
Yet, because the main conclusion persists for a wide range of the principal dynamics/geometry parameters, it strongly suggests that the partonic energy loss has a sizable collisional component.
\\[3mm]
{\bf Acknowledgments:} This work was supported by BMBF.

\end{document}